\documentclass[10pt,conference]{IEEEtran}

\usepackage[top = 0.73 in, bottom = 0.99 in, left = 0.63 in, right = 0.63 in]{geometry}
\usepackage{amsmath,amsfonts,amsthm}
\usepackage{algorithmic}
\usepackage{algorithm}
\usepackage{array}
\usepackage{textcomp}
\usepackage{stfloats}
\usepackage{url}
\usepackage{color}
\usepackage{verbatim}
\usepackage{graphicx}
\usepackage{cite}
\usepackage{tikz}
\usepackage{tikzscale}
\usepackage{circuitikz}
\usetikzlibrary{arrows.meta}
\usepackage{subfigure}
\usepackage[acronyms,nonumberlist]{glossaries}
\usepackage{nicefrac}

\newacronym{BS}{BS}{base station}
\newacronym[plural=UEs,
            longplural={user equipment}]{UE}{UE}{user equipment}
\newacronym{UL}{UL}{uplink}
\newacronym{DL}{DL}{downlink}
\newacronym{NOMA}{NOMA}{non-orthogonal multiple access}
\newacronym{OMA}{OMA}{orthogonal multiple access}
\newacronym{PASS}{PASS}{pinching antenna system}
\newacronym{PA}{PA}{pinching antenna}
\newacronym{LoS}{LoS}{line-of-sight}
\newacronym{SIC}{SIC}{successive interference cancellation}
\newacronym{BER}{BER}{bit error rate}
\newacronym{QPSK}{QPSK}{quadrature phase-shift keying}
\newacronym{AWGN}{AWGN}{additive white Gaussian noise}
\newacronym{QAM}{QAM}{quadrature amplitude modulation}
\newacronym{CF}{CF}{characteristic function}
\newacronym{PD}{PD}{power-domain}

\IEEEoverridecommandlockouts
\begin{document}

\title{BER Analysis and Optimization of Pinching-Antenna-Based NOMA Communications

\author{

\IEEEauthorblockN{Mahmoud AlaaEldin, Amy S. Inwood, Xidong Mu,~and Michail Matthaiou}

\IEEEauthorblockA{Centre for Wireless Innovation (CWI), Queen’s University Belfast, Belfast BT3 9DT, U.K.}

\IEEEauthorblockA {Emails: \{m.alaaeldin, a.inwood, x.mu, m.matthaiou\}@qub.ac.uk}
}

\thanks{This work was supported by the U.K. Engineering and Physical Sciences Research Council (EPSRC) grant (EP/X04047X/2) for TITAN Telecoms Hub. The work of M. Matthaiou was supported by the European Research Council (ERC) under the European Union’s Horizon 2020 research and innovation programme (grant agreement No. 101001331).}
}

\maketitle

\begin{abstract}
This paper presents the first bit error rate (BER) analysis of a pinching-antenna (PA)-based non-orthogonal multiple access (NOMA) communication system. The PA is assumed to be able to be placed anywhere along the waveguide and serves two NOMA user equipment (UEs) in both uplink (UL) and downlink (DL) scenarios. Exact closed-form expressions for the average BER of each user are derived under practical imperfect successive interference cancellation (SIC). These expressions are then used to optimize the PA location for minimizing the overall average BER of both UEs. In the UL case, the interference between the users’ channels introduces phase-dependent fluctuations in the BER cost function, making it highly non-convex with many local extrema. To address this challenge, a smoothing technique is applied to extract the lower envelope of the BER function, effectively suppressing ripples and enabling a reliable identification of the global minimum. In the DL case, a joint optimization of the PA location and NOMA power allocation coefficients is proposed to minimize the average BER. Simulation results verify the accuracy of the analytical derivations and the effectiveness of the proposed optimization methods. Notably, the UL results demonstrate that an optimally positioned PA can create the required received power difference between two equally powered UEs for reliable power-domain NOMA decoding under imperfect SIC.
\end{abstract}



\section{Introduction}
As applications, such as artificial intelligence, high-definition streaming, and augmented reality become commonplace, the demand for mobile connectivity is growing rapidly. Consequently, the demand for higher capacity and more efficient networks continues to intensify. One way to increase capacity is to increase the carrier frequency. However, this introduces greater propagation challenges, including stronger attenuation, stronger path loss, and higher sensitivity to blockages \cite{rangan_millimeter_2014}, making reliable access to \gls{LoS} paths critical.

\Glspl{PASS} are a recently proposed class of flexible antenna technologies, in which a dielectric particle, referred to as a \gls{PA}, is placed on a dielectric waveguide to radiate wireless signals \cite{fukuda_pinching_2022, ding_flexible_2025}. Spanning lengths in the order of meters, the dielectric waveguide provides the flexibility to reposition the \gls{PA} over a large range, creating a \gls{LoS} transceiver link. Note that \glspl{PASS} differ from other flexible-antenna technologies, such as fluid antennas and movable antennas, where adjustments are limited to the wavelength scale \cite{ding_flexible_2025}.
While \glspl{PASS} enable reliable communications at high carrier frequencies, spectrum resources remain limited and must be used efficiently. Each waveguide in a \gls{PASS} conveys an identical signal. Therefore, using \gls{PD}-\gls{NOMA} with \gls{SIC}, which allows all \glspl{UE} to simultaneously share the same resource block, in conjunction with \glspl{PASS} can offer improved spectral efficiency.

Existing studies on the synergy between \glspl{PASS} and \gls{NOMA} have primarily focused on analyzing and optimizing system-level metrics, such as the sum rate \cite{zhou_sum_2025,xu_rate_2025,lv_beam_2025}. However, there has been limited work om symbol-level metrics, which provide crucial insights into system reliability.

Therefore, in this work, we provide the first study on the \gls{BER} of \gls{PA}-assisted \gls{NOMA}, as well as the first study on \gls{PA}-assisted \gls{NOMA} with imperfect \gls{SIC} decoding. We consider a fundamental two-user single-antenna \gls{PASS}, where the \gls{PA} can be located at any point on the waveguide, and investigate both \gls{UL} and \gls{DL} scenarios. Exact \gls{BER} expressions are derived for the two users in the \gls{UL} \gls{PD}-\gls{NOMA} scenario employing \gls{QPSK} modulation and considering imperfect \gls{SIC} decoding. We then optimize the \gls{PA} position to minimize the derived average \gls{UL} \gls{BER}. Finally, we propose a method to jointly optimize the \gls{PA} position and power allocation coefficients in the \gls{DL} \gls{PD}-\gls{NOMA} scenario employing general $M$-\gls{QAM} modulation to minimize the average \gls{BER} of the users under imperfect \gls{SIC}.

\textit{Notation}: The probability of event $A$ is denoted $\mathrm{Pr}(A)$; $\mathrm{Pr}(A\mid B)$ represents the probability of event $A$ conditioned on event $B$; $P_\mathrm{e}(\cdot)$ denotes the probability of error; $\Re(z)$ and $\Im(z)$ denote the real and imaginary components of complex number $z$, respectively; $\angle z$ is the angle of complex number $z$; $\mathcal{CN}(\mu,\sigma^2)$ represents a complex Gaussian distribution with mean $\mu$ and variance $\sigma^2$; $Q(\cdot)$ is the Gaussian Q-function, and $\mathrm{sgn}(\cdot)$ is the signum function.


\section{System Model}
\label{sec:sysmodel}
\begin{figure}[ht]
    \centering
    \resizebox{0.485\textwidth}{!}{\includegraphics{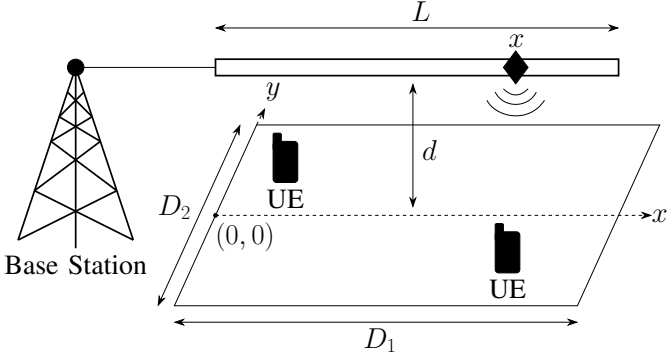}}
    \vspace{-0.1in}
    \caption{A \gls{PA}-based \gls{NOMA} communication system with two users.}
    \vspace{-0.1in}
    \label{fig:sys_model}
\end{figure}

Consider the \gls{PASS} shown in Fig. \ref{fig:sys_model}, where a single-antenna \gls{BS} serves two single-antenna \glspl{UE} via a single \gls{PA}. The \gls{PA} can be activated at any point $x$ on the waveguide of length $L$. The \glspl{UE} are distributed within a $D_1~\times~D_2$ $\mathrm{m}^2$ region, the \gls{BS} is located at $(x,y)=(0,0)$, and the waveguide is aligned along the $x$-axis with a vertical offset of $d$ m.

As the position of the \gls{PA} can be chosen to guarantee \gls{LoS}, and given that \glspl{PA} are primarily intended for high-frequency deployments, the channels between the \gls{PA} and the \glspl{UE} are assumed to be purely \gls{LoS}. Accordingly, the channel between \gls{UE} $k$ and the \gls{PA} located at position $x$ along the waveguide is modeled using the spherical-wave channel model \cite{zhang_beam_2022, wang_antenna_2025}: 
\begin{equation}  \label{sphr_ch_mod}
    h_k(x) = \frac{\eta\,\mathrm{e}^{-j\frac{2\pi}{\lambda}\left|\boldsymbol{\psi}_k-\boldsymbol{\psi}_p(x)\right|}}{\left|\boldsymbol{\psi}_k-\boldsymbol{\psi}_p(x)\right|}, 
\end{equation}
where $\boldsymbol{\psi}_k=(x_k,y_k,0)$ is the location of UE $k$, $\boldsymbol{\psi}_p(x)=(x,0,d)$ is the location of the \gls{PA} as a function of $x$, $\eta=\frac{c}{4\pi f_c}$, $c$ is the speed of light in a vacuum, $f_c$ is the carrier frequency and $\lambda$ is the wavelength. Additionally, we model the loss from the signal traveling within the waveguide as \cite{ding_flexible_2025}
\begin{equation}
    P_\mathrm{L}(x) = 10^{-\frac{\kappa}{20}\left|\boldsymbol{\psi}_p(0) - \boldsymbol{\psi}_p(x)\right|}\mathrm{e}^{-j\frac{2\pi}{\lambda_\mathrm{g}}\left|\boldsymbol{\psi}_p(0) - \boldsymbol{\psi}_p(x)\right|}, 
\end{equation}
where $\kappa$ is the average attenuation factor along the dielectric waveguide in dB/m, $\lambda_\mathrm{g}=\frac{\lambda}{n_\mathrm{eff}}$ is the guided wavelength, and $n_\mathrm{eff}$ is the effective refractive index of the dielectric waveguide. Therefore, the channel between UE $k$ and the BS via a \gls{PA} located at $x$ is 
\begin{equation}
    \tilde{h}_k(x) = P_\mathrm{L}(x) h_k(x). 
    \label{eq:htilde}
\end{equation}


\subsection{Uplink scenario}

In the \gls{UL} scenario, each \gls{UE} transmits its own signal, both of which are received by the \gls{PA}. Thus, the total superimposed received \gls{NOMA} signal at the \gls{BS} can be written as 
\begin{equation}
y^\mathrm{UL} = \sqrt{P_1}\tilde{h}_1(x)s_1 + \sqrt{P_2}\tilde{h}_2(x)s_2 + n, \label{eq:yUL} 
\end{equation}
where $P_1$ and $P_2$ are the transmit powers of \glspl{UE} 1 and 2, respectively, $s_1$ and $s_2$ are the signals corresponding to \glspl{UE} 1 and 2, respectively, while $n\sim\mathcal{CN}(0,2\sigma^2)$ is the \gls{AWGN}. In the \gls{UL} scenario, we assume that both \glspl{UE} employ gray-coded \gls{QPSK} modulation, such that $s_k$ is drawn from the constellation $\mathcal{S} = \{1+j, 1-j, -1-j, -1+j\}$, corresponding respectively to the bit pairs $(b_{k,1},b_{k,2}) = (0,1), (0,0), (1,0), (1,1)$.


\subsection{Downlink scenario}

In the \gls{DL} scenario, the signals intended for different \glspl{UE} are transmitted via a single radio-frequency chain. Therefore, they must be superimposed at the \gls{BS} using different power coefficients prior to transmission. Hence, the received \gls{NOMA} signal at the $k$-th \gls{UE} can be given as
\begin{equation}
y^\mathrm{DL}_k = \tilde{h}_k(x)\,s + n_k, \quad \forall k \in \{1, 2\},
\end{equation}
where 
\begin{equation}
s = \sqrt{P_\mathrm{T}} \left( \sqrt{\tfrac{\alpha}{\nu_1}} s_1 + \sqrt{\tfrac{1-\alpha}{\nu_2}} s_2 \right),
\end{equation}
is the superimposed signal transmitted by the \gls{BS}, $P_\mathrm{T}$ is the transmit power of the \gls{PA}, and $\alpha$ is the power allocation coefficient. The modulation symbol $s_k = s_{k,\mathrm{I}} + j s_{k,\mathrm{Q}}$ is drawn from a square \gls{QAM} alphabet, where the in-phase and quadrature components of $s_k$ can take values from the set $\{\pm 1, \pm 3, \dots, \pm \sqrt{M_k}-1\}$, while $M_k$ is the modulation order of \gls{UE} $k$. The scaling factor $\nu_k=\frac{2}{3}\left({M_k-1}\right)$ is used to normalize the power of $s_k$ to unity.


\section{Uplink and Downlink BER Analysis}

In this section, we present exact \gls{BER} expressions for each \gls{UE} in both the \gls{UL} and \gls{DL} scenarios by detailing the imperfect \gls{SIC} detection process. 


\subsection{Uplink scenario}

\subsubsection{BER Analysis of UE 1}
Without loss of generality, it is assumed that the received signal from \gls{UE} 1 has a higher amplitude than that from \gls{UE} 2 and is therefore decoded first, with the signal from \gls{UE} 2 treated as interference. To decode $s_1$, the \gls{BS} multiplies the total received signal in \eqref{eq:yUL} by $\mathrm{e}^{-j\angle \tilde{h}_1(x)}$ to remove the phase shift of the channel, leading to
\begin{equation}
    y^\mathrm{UL}_1{=} \sqrt{\!P_1}|\tilde{h}_1(x)|s_1 + \sqrt{\!P_2}\tilde{h}_2(x)\mathrm{e}^{-j\angle \tilde{h}_1(x)}s_2 + n\mathrm{e}^{-j\angle \tilde{h}_1(x)}\!, \notag
\end{equation}
which aligns $|\tilde{h}_1(x)|s_1$ with the \gls{QPSK} constellation. The rotated received signal from \gls{UE} 2 is treated as additional noise. The detection rule of $s_1$ can now be given as
\begin{equation}
    \hat{s}_1=\hat{s}_{1,\mathrm{I}} + j\hat{s}_{1,\mathrm{Q}}=\mathrm{sgn}(\Re(y^\mathrm{UL}_1)) +j\,\mathrm{sgn}(\Im(y^\mathrm{UL}_1)).
\end{equation}
The unconditional \gls{BER} of \gls{UE} 1 can be expressed as the average of the conditional \glspl{BER} over all possible pairs of transmitted \gls{QPSK} symbols $(s_1,s_2)$, such that
\begin{equation}
\mathrm{BER^\mathrm{UL}_{1}} = \tfrac{1}{4} \sum\limits_{s_1\in\mathcal{S}} \tfrac{1}{4} \sum\limits_{s_2\in\mathcal{S}} \mathrm{BER}_{1 \mid s_1, s_2}^{\mathrm{UL}}.
\end{equation}
The conditional \gls{BER}, given a transmitted symbol pair $(s_1,s_2)$, can be expressed as the average of the detection error probabilities of its two individual bits, such that
\begin{equation}
\mathrm{BER}_{1\mid s_1,s_2}^{\mathrm{UL}} = \tfrac{1}{2} \left( P_\mathrm{e}\left(b_{1,1} \mid s_1, s_2\right) + P_\mathrm{e}\left(b_{1,2} \mid s_1, s_2\right) \right), \label{eq:Pe_UL}
\end{equation}
where $b_{1,1}$ and $b_{1,2}$ are the bits transmitted by \gls{UE} 1, which are represented by the in-phase and quadrature components of $s_1$, $s_{1,\mathrm{I}}$ and $s_{1,\mathrm{Q}}$, respectively. The error probability of $b_{1,1}$ and $b_{1,2}$ is the likelihood that the Gaussian noise drives the received signal beyond the decision threshold of zero as
\begin{align}
P_\mathrm{e}\left( b_{1,1} {\mid} s_1, s_2 \right) &{=} \mathrm{Pr} \left(s_{1,\mathrm{I}}\, \Re(y^\mathrm{UL}_1) < 0 \right) = Q\left(\frac{\mu_{b_{1,1}}}{\sigma}\right), \label{b11_error}  \\ 
P_\mathrm{e}\left( b_{1,2} {\mid} s_1, s_2 \right) &{=} \mathrm{Pr} \left(s_{1,\mathrm{Q}}\, \Im(y^\mathrm{UL}_1) < 0\right) = Q\left(\frac{\mu_{b_{1,2}}}{\sigma}\right),  \label{b12_error}
\end{align}
where
\begin{align}
\mu_{b_{1,1}} & = \sqrt{\!P_1} |\tilde{h}_1(x)| \!+\! \sqrt{\!P_2} \, s_{1,\mathrm{I}} \, \Re\! \left( \tilde{h}_{2}(x) \mathrm{e}^{-j\angle \tilde{h}_1(x)} s_2 \right), \notag \\
\mu_{b_{1,2}} & = \sqrt{\!P_1} |\tilde{h}_1(x)| \!+\! \sqrt{\!P_2} \, s_{1,\mathrm{Q}} \, \Im \! \left( \tilde{h}_{2}(x) \mathrm{e}^{-j\angle \tilde{h}_1(x)} s_2 \right). \notag
\end{align}

\subsubsection{BER Analysis of UE 2}
\begin{figure}[!ht]
\centering
\vspace{-0.1 in}
\resizebox{0.485\textwidth}{!}{\includegraphics{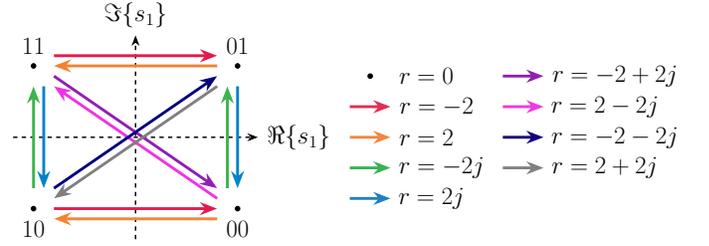}}
\caption{Residual interference vectors $r = s_1 - \hat{s}_1$, drawn from the transmitted symbol $s_1$ (origin) to the detected symbol $\hat{s}_1$ (endpoint).}
\vspace{-0.2in}
\label{fig:constellation}
\end{figure}

To decode $s_2$, the detected signal of \gls{UE} 1, $\sqrt{P}_1 \tilde{h}_1(x) \hat{s}_1$, is first subtracted from the received signal in \eqref{eq:yUL}. The residual is then multiplied by $\mathrm{e}^{-j\angle \tilde{h}_2(x)}$, which aligns the desired \gls{UE} 2 component with the \gls{QPSK} constellation, yielding 
\begin{multline}
    y_2^\mathrm{UL}= \sqrt{\!P_2}|\tilde{h}_2(x)|s_2 + \sqrt{\!P_1}\tilde{h}_1(x) \\ \times \mathrm{e}^{-j\angle \tilde{h}_2(x)}(s_1-\hat{s}_1) + n\mathrm{e}^{-j\angle \tilde{h}_2(x)}. \notag  
\end{multline}
Again, treating the non-desired signal as additional noise, the received signal from \gls{UE} 2 can be decoded as 
\begin{equation}
    \hat{s}_2=\mathrm{sgn}(\Re(y^\mathrm{UL}_{2})) +j\,\mathrm{sgn}(\Im(y^\mathrm{UL}_{2})). 
\end{equation}
To determine the unconditional \gls{BER} of \gls{UE} 2, all possible transmitted symbols $s_1$ and $s_2$, as well as all possible detected symbols $\hat{s}_1$ must be considered. For each transmitted pair $(s_1,s_2)$, $\hat{s}_1$ may take any QPSK value, with probability $\mathrm{Pr}(\hat{s}_1 {=} c \mid s_1,s_2), \forall c \in \mathcal{S}$. After decoding $\hat{s}_1$, the contribution of \gls{UE}~1 is subtracted via \gls{SIC}, leaving a residual interference of $R = s_1 - \hat{s}_1$ that may affect the detection of $s_2$. The conditional \gls{BER} for \gls{UE}~2 given this outcome is $\mathrm{BER}_{2 \mid s_2, R}^{\mathrm{UL}}$. 
By the law of total probability, the unconditional \gls{BER} of \gls{UE}~2 is obtained by summing the conditional \gls{BER} over all values of $R$, weighted by $\Pr(R = r \mid s_2)$, and subsequently averaging over all possible realizations of $s_2$, yielding
\begin{equation}
\mathrm{BER}^\mathrm{UL}_2 \!=\! \frac{1}{4} \! \sum_{s_2\in\mathcal{S}} \sum_{r \in \mathcal{R}} \mathrm{Pr}(R = r\mid s_2)  \mathrm{BER}_{2 \mid s_2, R}^{\mathrm{UL}}.
\end{equation}
The values that $R$ can take are determined by the $\hat{s}_1$ outcomes of the detection process of $s_1$. Each of the $4$ possibilities of $s_1$ being transmitted has $4$ possible decoded symbols, giving $16$ outcomes in total, some of which produce the same residual due to symmetry. The $9$ possible distinct residual values are
\begin{equation}
    R=r, \quad r \in \mathcal{R}=\{0,\pm2, \pm2j, \pm2\pm2j\},
\end{equation}
where $r=0$ corresponds to correct detection, $r=\pm 2$ or $\pm 2j$ to a single-bit error, and $r=\pm 2 \pm 2j$ to errors in both bits. These values are illustrated in Fig \ref{fig:constellation}.

To calculate the probability of $R=r$ given $s_2$, let us define
\begin{equation} \label{s1_hat_c}
\mathrm{Pr} (R = r \mid s_1,s_2) = \mathrm{Pr} (\hat{s}_1 = c: s_1 - c = r \mid s_1, s_2),
\end{equation}
where $r \in \mathcal{R}$ and $c \in \mathcal{S}$. The probability of a given residual is the sum of the probabilities of all $(s_1, \hat{s}_1)$ pairs that lead to that residual. As all $s_1$ are equally likely, the explicit dependence on $s_1$ can be removed by averaging over it, such that 
\begin{equation}
\mathrm{Pr} (R = r \mid s_2) = \frac{1}{4} \sum\limits_{s_1 \in \mathcal{S}} \mathrm{Pr} (R = r \mid s_1, s_2). 
\end{equation}
To evaluate $\mathrm{BER}_2^\mathrm{UL}$, expressions for $\mathrm{Pr}(\hat{s}_1 = c \mid s_1,s_2)$ and $\mathrm{BER}_{2 \mid s_2, R}^{\mathrm{UL}}$ are required. As the Gaussian noise is circularly symmetric, and the in-phase and quadrature components are independent, $\mathrm{Pr}(\hat{s}_1 = c \mid s_1,s_2)$ can be expressed as 
\begin{equation}
\mathrm{Pr}(\hat{s}_1 {=} c \mid s_1,s_2) \!=\! \mathrm{Pr}(\hat{s}_{1,\mathrm{I}} {=} c_{\mathrm{I}} \mid s_1, s_2) \mathrm{Pr}(\hat{s}_{1,\mathrm{Q}} {=} c_{\mathrm{Q}} \mid s_1, s_2).\notag
\end{equation}
This reduces the two-dimensional QPSK decision rule to two independent one-dimensional detection problems along the in-phase and quadrature axes. The noise-free signal component along each axis can be defined as 
\begin{align}
    \mu_{\hat{s}_{1,\mathrm{I}}} &\!= \!\sqrt{P_1}|\tilde{h}_{1}(x)|s_{1,\mathrm{I}}+ \sqrt{P_2}\,\Re\left(\tilde{h}_2(x)\mathrm{e}^{-j\angle\tilde{h}_{1}(x)}s_2\right)\!, \notag\\
    \mu_{\hat{s}_{1,\mathrm{Q}}} & \!= \!\sqrt{P_1}|\tilde{h}_{1}(x)|s_{1,\mathrm{Q}}+ \sqrt{P_2}\,\Im\left(\tilde{h}_2(x)\mathrm{e}^{-j\angle\tilde{h}_{1}(x)}s_2\right)\!. \notag
\end{align}
Therefore, the detection probabilities along each axis are
\begin{equation}
\mathrm{Pr}(\hat{s}_{1,\mathrm{I}} {=} c_{\mathrm{I}} {\mid} s_1,s_2) {=} \mathrm{Pr}(-c_{\mathrm{I}} \, \Re(y^\mathrm{UL}_2) {<} 0) {=}  Q\!\left(\!-c_{\mathrm{I}} \frac{ \mu_{\hat{s}_{1,\mathrm{I}}} } {\sigma} \!\right)\!\!, \label{eq:probshat1I}
\end{equation}
\begin{equation}
     \!\!\!\mathrm{Pr}(\hat{s}_{1,\mathrm{Q}} {=} c_{\mathrm{Q}} {\mid} s_1\!,\!s_2) {=} \mathrm{Pr}(-c_{\mathrm{Q}} \, \!\Im(y^\mathrm{UL}_2) {<} 0) {=} Q\!\left( \! -c_{\mathrm{Q}} \frac{ \mu_{\hat{s}_{1,\mathrm{Q}}} } {\sigma} \!\!\right)\!\!.\!\! \label{eq:probshat1Q}
\end{equation}
Note that the negative signs arise as \eqref{eq:probshat1I} and \eqref{eq:probshat1Q} correspond to detection probabilities rather than error probabilities. Finally, following the same process as for $\mathrm{BER}_{1 \mid s_1, s_2}^{\mathrm{UL}}$,
\begin{equation}
\mathrm{BER}_{2 \mid s_2, R}^{\mathrm{UL}} = \tfrac{1}{2} \left( P_\mathrm{e} \left( b_{2,1} \mid s_2, R \right) + P_\mathrm{e} \left( b_{2,2} \mid s_2, R \right) \right), \label{ber_UE2}
\end{equation}
where $b_{2,1}$ and $b_{2,2}$ are the bits transmitted by \gls{UE} 2, which are represented by the in-phase and quadrature components of $s_2$, $s_{2,\mathrm{I}}$ and $s_{2,\mathrm{Q}}$, respectively. The conditional error probability of $b_{2,1}$ and $b_{2,2}$ in (\ref{ber_UE2}) can be given as 
\begin{align}
P_\mathrm{e} \big( b_{2,1} \mid s_2, R \big) &= \left(s_{2,\mathrm{I}}\, \Re(y^\mathrm{UL}_2) < 0 \right) = Q\left( \frac{\mu_{s_{2,\mathrm{I}}}}{\sigma} \right),\\
P_\mathrm{e} \big( b_{2,2} \mid s_2, R \big) &= \left(s_{2,\mathrm{Q}}\, \Re(y^\mathrm{UL}_2) < 0 \right) = Q\left( \frac{\mu_{s_{2,\mathrm{Q}}}}{\sigma} \right),
\end{align}
and \vspace{-0.05 in}
\begin{align}
    \mu_{s_{2,\mathrm{I}}} &= \sqrt{P_2}|\tilde{h}_2(x)| + \sqrt{P_1}s_{2,\mathrm{I}}\,\Re\left(\tilde{h}_1(x)\mathrm{e}^{-j\angle\tilde{h}_{2}(x)}R\right), \notag \\
    \mu_{s_{2,\mathrm{Q}}} &= \sqrt{P_2}|\tilde{h}_2(x)|+ \sqrt{P_1}s_{2,\mathrm{Q}}\,\Im\left(\tilde{h}_1(x)\mathrm{e}^{-j\angle\tilde{h}_{2}(x)}R\right). \notag
\end{align}


\subsection{Downlink scenario}
In the \gls{DL}, each \gls{UE} receives both transmitted signals multiplied by the same channel coefficient. Therefore, applying a phase shift of $\mathrm{e}^{-j\angle \tilde{h}_k(x)}$ at \gls{UE} $k$ to align $\tilde{h}_k(x)s$ with the \gls{QAM} constellation,
\begin{equation}
    \bar{y}^\mathrm{DL}_k = |\tilde{h}_k(x)|s + n_k\mathrm{e}^{-j\angle \tilde{h}_k(x)}.
\end{equation}
The unconditional average BER of \gls{UE} $k$ was derived in \cite{alaaeldin3} as
\begin{equation}  \label{ber_express_dl}
\mathrm{BER}_{k}^{\mathrm{DL}}\!\!=\!\! \sum_{q=1}^{N_Q} \!c_q  Q\!\!\left(\!\! \frac{ \sqrt{\!P_\mathrm{T}} |\tilde{h}_k(x)|\!  \Big(\!a_{1,q} \sqrt{\!\frac{\alpha}{\nu_1}} \!+\! a_{2,q} \sqrt{\frac{\!1 {-} \alpha}{\nu_2}} \Big) }{\sigma}\!\!\right)\!\!,\!
\end{equation}
where $N_Q$ is the total number of $Q(\cdot)$ functions in the conditional BER expression of \gls{UE} $k$, and $a_{1,q}$, $a_{2,q}$ and $c_{q}$ are constants that depend on the modulation order and can be calculated using \cite[Algorithm 1]{alaaeldin3}.


\section{BER minimization-based optimization of the PA-based NOMA system}

In this section, we propose optimization schemes to minimize the average \gls{BER} of the two \glspl{UE} in the \gls{UL} and \gls{DL}.

\subsection{Optimization of the UL scenario}
\label{sec:optUL}

In the \gls{UL} scenario, the location of the \gls{PA} can be adjusted to create a sufficient difference in the received power from the 2 \glspl{UE} at the \gls{BS} and thus facilitate successful \gls{SIC}-based \gls{PD}-\gls{NOMA} decoding. To achieve this, the \gls{PA} can be positioned closer to the stronger \gls{UE} (in this case, $\mathrm{UE}$~1, without loss of generality), which is the first user in the \gls{SIC} decoding order. This placement reduces the path loss associated with the stronger \gls{UE} and ensures that its received signal at the \gls{BS} has a higher power level than that of the weaker user (in this case, $\mathrm{UE}$~2). This \gls{PA} placement reduces the decoding error of $s_1$ (where $s_2$ acts as interference) and thereby mitigates the error propagation caused by imperfect \gls{SIC} when subtracting $\sqrt{P_1}\tilde{h}_1(x)\hat{s}_1$ from the received signal $y^{\mathrm{UL}}$ to decode $s_2$.

As the derived expressions of $\mathrm{BER}_1^{\mathrm{UL}}$ and $\mathrm{BER}_2^{\mathrm{UL}}$ are functions of the \gls{PA} location, $x$ can be optimized to minimize the average $\mathrm{BER}$ of the two users. Therefore, the optimization problem of $x$ can be formulated as
\begin{subequations} \label{non_smooth_prob}
\begin{align}
\min_{x} \quad & f(x) = 10 \log_{10} \big( \mathrm{BER}_1^{\mathrm{UL}} (x) + \mathrm{BER}_2^{\mathrm{UL}} (x)  \big),  \label{cost_fun}  \\ 
{\mathrm {s.t.}} \quad & 0 \leq x \leq L,
\end{align}
\end{subequations}
where the $\log$ transformation in (\ref{cost_fun}) is made to pronounce the structure of the objective function and ease the numerical search process. While the cost function has a clear trend, it rapidly fluctuates (as seen in Fig. \ref{fig:opt_pos} in Sec. \ref{subsec:numres_UL}). These fluctuations are a result of the superposition of $h_1(x)$ and $h_2(x)$ at the \gls{PA}. Each channel has a different phase shift, which causes constructive interference (local maxima) or destructive interference (local minima) with varying $x$ as evident in (\ref{sphr_ch_mod}).

Due to these fluctuations, directly optimizing $f(x)$ with numerical methods, such as gradient descent, is impractical, as it quickly converges to local minima. Therefore, the moving minimum sliding window technique is applied to produce a smooth version of $f(x)$ for minimization. This technique computes the minimum value of $f(x)$ within a local sliding window. For a discrete version of the function, $f(nT)$, where $T$ denotes the sampling period, the operation is defined as 
\begin{equation}  \label{smoothing}
    \bar{f}(x) = \min_{n \in W_x} f(nT),
\end{equation}
where $W_x$ denotes the set of indices within a window centered at $x$. This nonlinear filtering process is equivalent to a morphological erosion, effectively suppressing short-term peaks while preserving local minima. When applied to a non-smooth or noisy function, the resulting $\bar{f}(x)$ provides a locally adaptive estimate of the lower envelope, yielding a curve that captures the underlying minimum structure of the function. To accurately capture the phase-induced variations of $f(x)$, $T$ should be less than $\lambda$. Moreover, the width of the sliding window should span multiple wavelengths to ensure sufficient averaging of the fluctuations in $f(x)$. A gradient-based method is then employed to minimize the resulting smooth and continuous function $\bar{f}(x)$ over the interval $0 \leq x \leq L$. Lastly, a fine-tuning stage is performed by evaluating $f(x)$ at $2N+1$ uniformly spaced points in the interval
$[x^\star_{\mathrm{sm}}-\Delta,\; x^\star_{\mathrm{sm}}+\Delta]$.
Here, $x^\star_{\mathrm{sm}}$ denotes the \gls{PA} position corresponding to the minimum of $\bar{f}(x)$,
$\Delta = N\delta$ and should be in the order of wavelengths, and $\delta \ll \lambda$ is the sample spacing.
The minimum sample is selected, which, as $N \to \infty$, converges to the global minimum.

\subsection{Optimization of the DL scenario}
\label{sec:DL_opt}

In this subsection, an average $\mathrm{BER}$ minimization-based joint optimization method is proposed to jointly optimize the \gls{PA} position $x$, and the power allocation coefficient $\alpha$. Unlike in the \gls{UL}, the optimization of the \gls{DL} scenario does not require any smoothing. The \gls{DL} $\mathrm{BER}$ expression in (\ref{ber_express_dl}) depends solely on the magnitude of the channel, $|\tilde{h}_k(x)|$, and therefore does not experience the rapid phase fluctuations observed in the \gls{UL} case. However, in the \gls{DL} scenario, both the \gls{PA} location, $x,$ and the power control coefficient, $\alpha$, shall be jointly optimized to minimize the average $\mathrm{BER}$ of the users. Unlike the \gls{UL} scenario, the \gls{PA} position alone is not enough to enable \gls{NOMA} - if the signals intended for both \glspl{UE} are transmitted with the same power, it will be impossible for each user to apply \gls{SIC} and successfully decode the symbols. Therefore, the optimization problem can be formulated as
\begin{subequations} \label{dl_opt_prob}
\begin{align}
\min_{x, \alpha} \quad & 10 \log_{10} \left( \mathrm{BER}_1^{\mathrm{DL}} (x, \alpha) + \mathrm{BER}_2^{\mathrm{DL}} (x, \alpha)  \right),  \label{cost_fun_dl}  \\ 
{\mathrm {s.t.}} \quad & 0 \leq x \leq L, \quad 0 \leq \alpha \leq 1.
\end{align}
\end{subequations}
The cost function in (\ref{cost_fun_dl}) is smooth and continuous in $x$ and $\alpha$ (as seen in Fig. \ref{fig:opt_pos_DL} of Sec. \ref{subsec:numres_DL}). Thus, (\ref{dl_opt_prob}) can be solved using a gradient-based method, as the cost function and constraints are continuous and have continuous first derivatives.\footnote{Note that the optimization procedure should be repeated with multiple random initializations to ensure convergence to the global minimum. This lies significantly below any local minimum, making it easily identifiable.}


\section{Numerical Results}
\label{sec:numresults}
This section verifies the analytical results and explores the
system behavior. We consider a \gls{PA} of length $L=20$ m at a height of $d=3$~m above a $D_1\times D_2 = 20\times4$~m$^2$ environment.  The attenuation factor along the dielectric is $\kappa=0.1$ dB/m and the effective refractive index of the dielectric is $\eta_\mathrm{eff}=1.4$ \cite{ding_flexible_2025}. We assume that \gls{UE} 1 is located at $(x,y)=(3,-1)$ and \gls{UE} 2 is located at $(x,y)=(18,3)$. In the \gls{DL} scenario, the \gls{BS} employs \gls{QPSK} modulation to the signal intended for \gls{UE} 1 and 16-\gls{QAM} to the signal intended for \gls{UE} 2. In the \gls{UL} case, it is assumed that both \glspl{UE} are transmitting at the same transmit power, that is, $P_1=P_2$. The carrier frequency is set to $f_c =28$~GHz, and the noise variance is $\sigma^2=-90$ dBm. For simulated results, $10^6$ symbols are generated.

\subsection{UL PA-Based NOMA}
\label{subsec:numres_UL}
\begin{figure}[ht]
\centering
\vspace{-0.1in}
\includegraphics[width=\columnwidth]{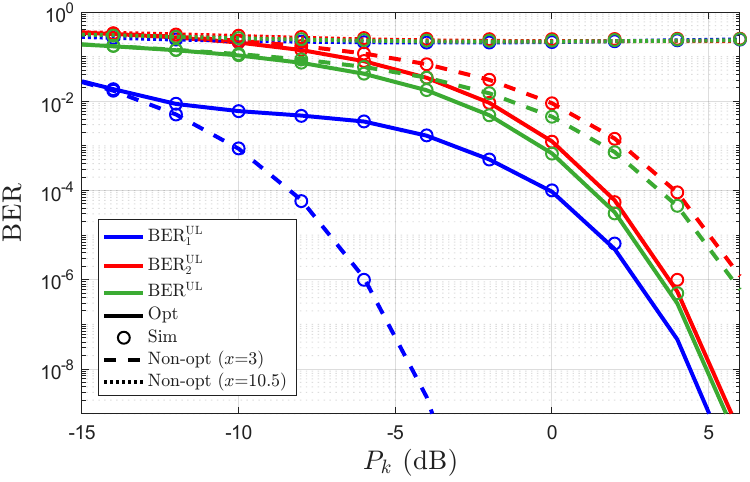}
\vspace{-0.25in}
\caption{\glspl{BER} of each \gls{UE} and the average \gls{BER} in the two user \gls{UL} \gls{PA}-assisted \gls{NOMA} system for both optimized and non-optimized PA placement.}
\vspace{-0.05in}
\label{fig:UL}
\end{figure}

Figure \ref{fig:UL} plots the \gls{BER} of each \gls{UE} and the average \gls{BER} against the transmit power for an \gls{UL} \gls{PA}-assisted \gls{NOMA} system. Results are shown for an optimally placed \gls{PA}, a \gls{PA} placed halfway between the \glspl{UE} at $x=10.5$, and a \gls{PA} placed next to \gls{UE} 1 at $x=3$. The sliding window applied for smoothing spans $20\lambda$, i.e., $20$ samples, and $T=0.01$ m.

It can be seen that there is excellent agreement between the analytical and simulated results for all scenarios considered. When the \gls{PA} is located at $x=10.5$, the \glspl{BER} exhibit a pronounced error floor. This is a consequence of the insufficient power difference between the \glspl{UE}, which prevents the effective operation of \gls{NOMA}. In contrast, placing the \gls{PA} at a location other than the midpoint between the two \glspl{UE} removes the error floor. Given the strong distance dependence of the channel coefficient in \eqref{eq:htilde}, the \gls{PA} can be strategically positioned to induce different path lengths between itself and the \glspl{UE}, providing the power difference required for successful \gls{NOMA}. These results demonstrate that a \gls{PA} can be leveraged to enable \gls{NOMA} and thus improve the utilization of limited spectrum.

To optimize the sum rate of a \gls{PA}-based \gls{NOMA} system, the best performance can often be obtained by maximizing the performance of the strongest \gls{UE} \cite{wang_antenna_2025,zhou_sum_2025}, which tends to involve placing the \gls{PA} near the strong \gls{UE}. However, as the focus in this work is the average \gls{BER}, it is necessary for both \glspl{UE} to achieve reliable error performance. Placing the \gls{PA} at $x=3$, the closest position to \gls{UE}~1, results in excellent \gls{BER} performance for \gls{UE}~1. However, the average \gls{BER} is worse than the optimized case because of the degraded performance of \gls{UE}~2. This shows that the optimization approach inherently ensures fairness.

\begin{figure}[ht]
\vspace{-0.1in}
\centering
\includegraphics[width=\columnwidth]{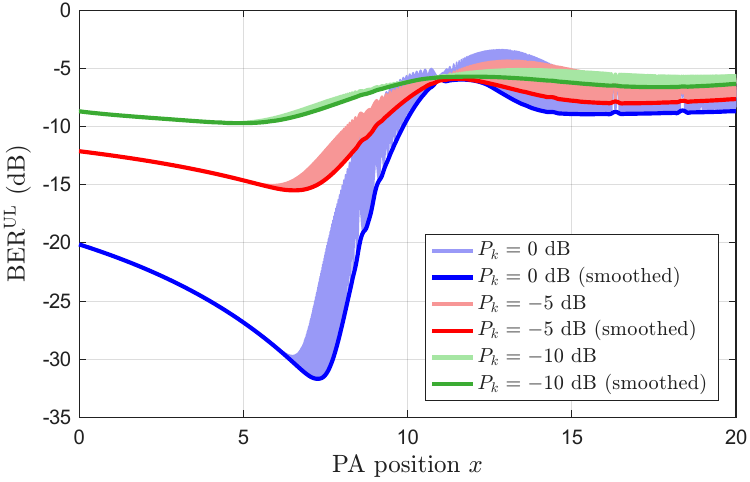}
\vspace{-0.25in}
\caption{Average \gls{UL} \gls{BER} in dB as a function of \gls{PA} position.}
\vspace{-0.05in}
\label{fig:opt_pos}
\end{figure}

Figure \ref{fig:opt_pos} plots the average \gls{UL} \gls{BER} as a function of \gls{PA} position for a range of \gls{UE} transmit powers, as well as the smoothed function used for the \gls{UL} optimization. 

The \gls{BER} function rapidly fluctuates with antenna position for all transmit power levels. The high carrier frequency means that even a position shift in the order of millimeters corresponds to a significant fraction of the wavelength, resulting in a large phase change. For example, in some regions of the dielectric in the $P_k=0$ dB scenario, a \gls{PA} position change of 1~mm corresponds to an average \gls{BER} change of around 10 dB. These oscillations demonstrate the necessity of smoothing the objective function before optimization as outlined in Sec.~\ref{sec:optUL}.

Additionally, at lower transmit power levels, the optimal \gls{PA} position shifts closer to \gls{UE}~1. This is because the required power disparity between the \glspl{UE} arises solely from their relative distances to the \gls{PA}. At lower power, a positional change yields a smaller absolute change in received power, necessitating a larger distance difference to achieve sufficient disparity. Conversely, at higher transmit power, lower average \glspl{BER} are attainable, but this requires both \glspl{UE} to maintain good error performance. In all cases, the highest average \gls{BER} occurs when the \gls{PA} is close to equidistant from both \glspl{UE}.

\subsection{DL PA-based NOMA}
\label{subsec:numres_DL}
\begin{figure}[ht]
\centering
\includegraphics[width=\columnwidth]{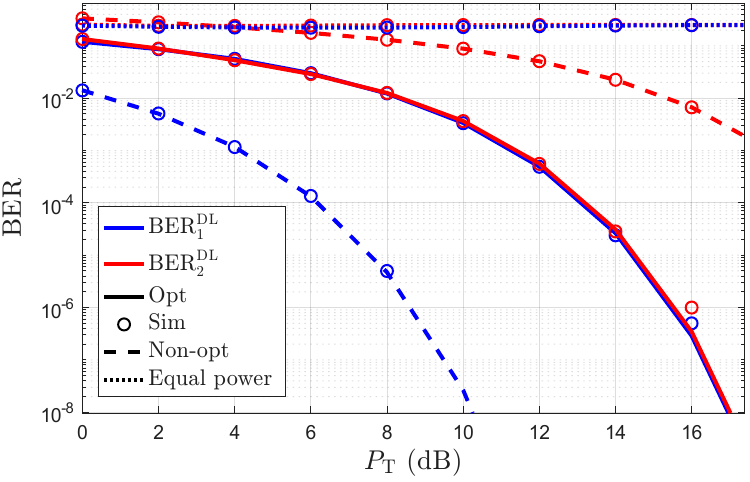}
\vspace{-0.25in}
\caption{\glspl{BER} of each \gls{UE} in a two-user \gls{DL} \gls{PA}-based \gls{NOMA} system for both optimized and non-optimized PA placement and power control coefficients.}
\label{fig:DL}
\vspace{-0.25in}
\end{figure}

Figure \ref{fig:DL} shows the \gls{BER} of the received signal at each \gls{UE} for a range of transmit powers for a \gls{DL} \gls{PA}-based \gls{NOMA} system. Results are shown for a system jointly optimized using the method detailed in Sec. \ref{sec:DL_opt}, a non-optimized system where $x=10$ and $\alpha=0.9$, and a system where the \gls{PA} position is optimized but $\alpha=0.5$, leading to the signals intended for each \gls{UE} being transmitted with equal power.

\begin{figure}[ht]
\centering
\includegraphics[width=\columnwidth]{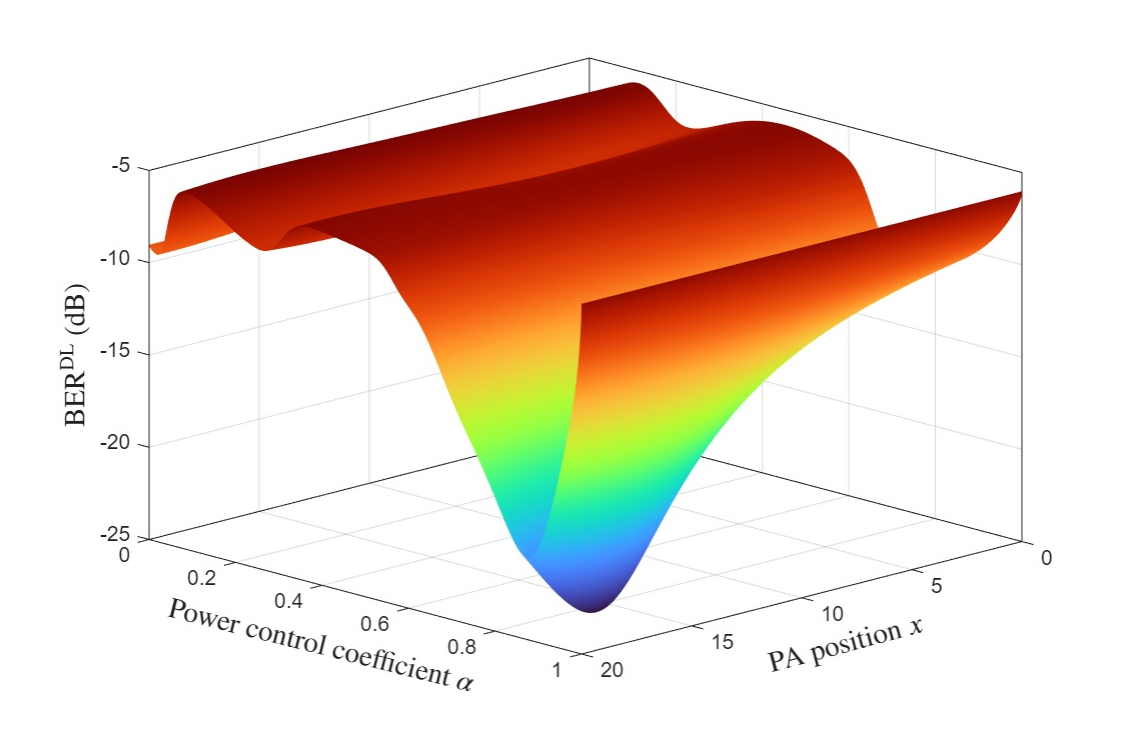}
\vspace{-0.3in}
\caption{Average \gls{DL} \gls{BER} in dB as a function of the \gls{PA} position and power allocation coefficient $\alpha$.}
\vspace{-0.16in}
\label{fig:opt_pos_DL}
\end{figure}

Again, it is evident that there is excellent agreement between simulated and analytical results. Similar to the \gls{UL} case, non-optimized parameters can result in the \gls{BER} of one \gls{UE} being significantly stronger than the other. This has a negative effect on the average \gls{BER}, which is dominated by the larger \gls{BER}. Optimization leads to equivalent \glspl{BER} for all \glspl{UE}. This same result would be achieved by minimizing the maximum individual \gls{BER}. However, jointly optimizing the parameters for the average \gls{BER} results in a simpler optimization problem, as there is only one function being optimized rather than multiple being solved iteratively. The \gls{BER} of both \glspl{UE} converging to the same value after optimization is possible due to the fact that the power allocation coefficient and \gls{PA} position are jointly optimized. In the \gls{UL} scenario, the results in Fig. \ref{fig:UL} show a gap between the optimized \gls{BER} of \glspl{UE} 1 and 2. In that scenario, both \glspl{UE} transmit with equal power. Therefore, the \gls{PA} position must create a sufficient power difference in the received signals from each \gls{UE} to enable \gls{NOMA}. This results in one \gls{UE} experiencing a stronger received signal and consequently achieving a lower \gls{BER}. As discussed in Sec. \ref{sec:DL_opt}, in the \gls{DL} case, it is not possible for \gls{PD}-\gls{NOMA} to function when the signals corresponding to each \gls{UE} are transmitted with the same power due to interfering \gls{QAM} constellations, making successful decoding at the receiver impossible. This is shown in Fig. \ref{fig:DL}, where the \glspl{BER} in the equal power scenario do not improve with higher transmit powers. Figure \ref{fig:opt_pos_DL} further details the impact of the relationship between the power control coefficient $\alpha$ and the \gls{PA} position on the \gls{BER}; when the power is shared equally, the average \gls{BER} is always high.

Figure \ref{fig:opt_pos_DL} also shows that the average \gls{DL} \gls{BER} function is smooth. This is in contrast to the average \gls{UL} \gls{BER} functions shown in Fig. \ref{fig:opt_pos}, which oscillate rapidly due to small changes in \gls{PA} position leading to large changes in phase. The smoothness in the \gls{DL} function is due to the unconditional \gls{BER} of each \gls{UE} only being a function of the magnitude of the channel $h_k(x)$. Moreover, unlike the \gls{UL} case, there are no fluctuations arising from the superposition of two channels with different phases. However, this is unique to systems with a single \gls{PA} - multiple \glspl{PA} will result in channels from each being superimposed at the \glspl{UE}, leading to constructive and destructive interference and thus fluctuations in the \gls{DL} average \gls{BER} function.


\section{Conclusion}
This paper provided the first \gls{BER} analysis of \gls{PA}-based \gls{PD}-\gls{NOMA} systems with the aid of practical imperfect \gls{SIC} decoding. We derived the exact average \gls{BER} for two-user \gls{UL} and \gls{DL} systems. The proposed \gls{PA} location optimization scheme for the \gls{UL} and the joint \gls{PA}–power allocation optimization scheme for the \gls{DL} were shown to effectively minimize the average \glspl{BER} of the two scenarios. Simulation results confirmed the accuracy of the analysis and demonstrated that the optimized \gls{PASS}-\gls{NOMA} system significantly outperforms the non-optimized counterpart. Both optimization methods inherently ensure fairness among \glspl{UE} and eliminate error floors. The results of the optimization schemes concluded that the optimized \gls{PA} location should be closer to the first user in the \gls{SIC} order in the \gls{UL} scenario, while the \gls{PA} should be placed closer to the second user in the \gls{SIC} order in the \gls{DL} scenario. These findings highlight the potential of strategic \gls{PA} placement to enable successful \gls{NOMA} under imperfect \gls{SIC} and improve the reliability and efficiency of \gls{PA}-based \gls{NOMA} systems.

\bibliographystyle{IEEEtran}
\bibliography{IEEEabrv, referencesPASS.bib}

\end{document}